\PassOptionsToPackage{unicode}{hyperref}
\PassOptionsToPackage{hyphens}{url}
\documentclass[
  11pt,
]{article}
\usepackage{amsmath,amssymb}
\usepackage{iftex}
\ifPDFTeX
  \usepackage[T1]{fontenc}
  \usepackage[utf8]{inputenc}
  \usepackage{textcomp} 
\else 
  \usepackage{unicode-math} 
  \defaultfontfeatures{Scale=MatchLowercase}
  \defaultfontfeatures[\rmfamily]{Ligatures=TeX,Scale=1}
\fi
\usepackage{lmodern}
\ifPDFTeX\else
\fi
\IfFileExists{upquote.sty}{\usepackage{upquote}}{}
\IfFileExists{microtype.sty}{
  \usepackage[]{microtype}
  \UseMicrotypeSet[protrusion]{basicmath} 
}{}
\usepackage{xcolor}
\usepackage{longtable,booktabs,array}
\usepackage{calc} 
\usepackage{etoolbox}
\makeatletter
\patchcmd\longtable{\par}{\if@noskipsec\mbox{}\fi\par}{}{}
\makeatother
\IfFileExists{footnotehyper.sty}{\usepackage{footnotehyper}}{\usepackage{footnote}}
\makesavenoteenv{longtable}
\usepackage{graphicx}
\makeatletter
\def\maxwidth{\ifdim\Gin@nat@width>\linewidth\linewidth\else\Gin@nat@width\fi}
\def\maxheight{\ifdim\Gin@nat@height>\textheight\textheight\else\Gin@nat@height\fi}
\makeatother
\setkeys{Gin}{width=\maxwidth,height=\maxheight,keepaspectratio}
\makeatletter
\def\fps@figure{htbp}
\makeatother
\usepackage{pos}
\usepackage[absolute,overlay]{textpos}

\author[a,b]{Michele Della Morte}
\author[a,b]{Benjamin J{\"a}ger}
\author*[a,b]{Sofie~Martins}
\author[c]{Justus Tobias Tsang} 
\author[d]{Felix P. G. Ziegler}

\affiliation[a]{IMADA, University of Southern Denmark, Campusvej 55, 5230 Odense M, Denmark}
\affiliation[b]{$\hbar$QTC, University of Southern Denmark, Campusvej 55, 5230 Odense M, Denmark}
\affiliation[c]{CERN, Theoretical Physics Department, 1211 Geneva 23, Switzerland}
\affiliation[d]{School of Physics and Astronomy, The University of Edinburgh, EH9 3FD Edinburgh, United Kingdom}

\emailAdd{dellamor@qtc.sdu.dk}
\emailAdd{jaeger@imada.sdu.dk}
\emailAdd{martinss@imada.sdu.dk}
\emailAdd{j.t.tsang@cern.ch}
\emailAdd{felix.ziegler@zsw-bw.de}

\abstract{We examine one-flavour $SU(N_c)$ gauge theories, where $N_c$ denotes the number of colors, with one fermion in the antisymmetric representation as a candidate to approximate $\mathcal{N}=1$ super Yang Mills due to their equivalence in the large-$N_c$ limit. Summarising results on spectral evaluations of $N_c=3$, we will report on the progress of dynamical calculations for $N_c>3$. We discuss cut-off effects and challenges in configuration generation.

\begin{textblock}{20}(15.0,1.70)
CERN-TH-2023-242
\end{textblock}%
}

\FullConference{%
	The 40th International Symposium on Lattice Field Theory (Lattice 2023),\\
	July 31st - August 4th, 2023\\
	Fermi National Accelerator Laboratory
	}

\usepackage{flafter}
\usepackage{float}
\usepackage{pos}
\usepackage{natbib}
\usepackage{subfigure}
\usepackage{booktabs}
\usepackage{longtable}
\usepackage{array}
\usepackage{multirow}
\usepackage{wrapfig}
\usepackage{colortbl}
\usepackage{pdflscape}
\usepackage{tabu}
\usepackage{threeparttable}
\usepackage{threeparttablex}
\usepackage[normalem]{ulem}
\usepackage{makecell}
\usepackage{xcolor}
\ifLuaTeX
  \usepackage{selnolig}  
\fi
\usepackage[]{natbib}
\nocite{*}
\IfFileExists{bookmark.sty}{\usepackage{bookmark}}{\usepackage{hyperref}}
\IfFileExists{xurl.sty}{\usepackage{xurl}}{} 
\hypersetup{
  pdftitle={Towards the super Yang-Mills spectrum at large N\_c},
  hidelinks,
  pdfcreator={LaTeX via pandoc}}

\title{Towards the super Yang-Mills spectrum at large \(N_c\)}
\date{\vspace{-2.5em}2023-07-19}

\begin{document}
\maketitle

{
\setcounter{tocdepth}{2}
\tableofcontents
}
\hypertarget{motivation}{%
\section{Motivation}\label{motivation}}

We are interested in studying the large-\(N_c\) limit of a one-flavour theory with one fermion in the two-index antisymmetric representation, following the proposal of \citep{CORRIGAN197973} to use this theory as a proxy to predict the low-lying mesonic spectrum of a supersymmetric theory. In order to test phenomenological predictions made in \citep{Sannino:2003xe} and \citep{Armoni:2005qr}, we have performed a lattice study for the case \(N_c=3\) published in \citep{DellaMorte:2023ylq}. We now wish to, on the one hand, produce this result for larger numbers of colours and, on the other hand, estimate the impact of cut-off effects on the result we have obtained so far. This is a progress report on this project.

\hypertarget{n_c3-summarized}{%
\section{\texorpdfstring{\(N_c=3\) summarized}{N\_c=3 summarized}}\label{n_c3-summarized}}

We are using the Symanzik-improved gauge action and one flavour of clover-improved Wilson fermions in the two-index antisymmetric representation at \(c_{\mathrm{sw}}=1\) with \texttt{OpenQCD} \citep{openqcd}. Due to having just a single flavour, we need the RHMC \citep{Clark:2006fx}.

Evaluating the spectrum of this one-flavour theory yields the low-lying mesonic states that are the supersymmetric analogon to the lowest-energy states of the glueball spectrum. We can obtain masses for these gluodynamic states by evaluating the connected and disconnected contributions to the correlation function and extracting the mass.

However, constructing the theory like this, we assume massive gluinos. In order to predict the actual realisation of massless gluinos, we need to take the chiral limit. As demonstrated in \citep{Francis:2018xjd}, the chiral point is characterized by vanishing topological susceptibility. Extrapolating to the point of vanishing topological susceptibility is equivalent to taking the mass in the connected contributions to zero. As in \citep{DellaMorte:2023ylq}, we denote the particle with this mass \(m_{\pi}^{\mathrm{fake}}\) as \textit{fake} pion. Taking the mass of the fake pion to zero is equivalent to taking the mass of the gluino constituents to zero.

Our \(N_c=3\) study focused on testing the phenomenological predictions \citep{Sannino:2003xe, Armoni:2005qr} for the ratio between pseudoscalar and scalar masses \(m_{\mathrm{P}} / m_{\mathrm{S}}\). This has the advantage that we can estimate the size of \(1/N_c\) effects by using \(N_c=3\) without a number of technical difficulties; for example, there is no necessity to set the scale.

The phenomenological prediction we aim to reproduce has been derived by \citep{Sannino:2003xe} and \citep{Armoni:2005qr} with different methods. Here, \citep{Sannino:2003xe} states the ratio between the scalar and pseudoscalar meson to be

\begin{equation}
\dfrac{m_{\mathrm{P}}}{m_{\mathrm{S}}} = 1 - \dfrac{22}{9N_c} - \dfrac{4}{9}\beta \lesssim 0.185    
\label{eq:estimate-sannino-shifman}
\end{equation}

up to unknown \(\mathcal{O}(1/N_c^2)\) effects, whereas \citep{Armoni:2005qr} arrives at a consistent result of

\begin{equation}
\dfrac{m_{\mathrm{P}}}{m_{\mathrm{S}}} = \dfrac{N_c-2}{N_c}\,.
\label{eq:estimate-armoni-imeroni}
\end{equation}

Note that \(\beta\) in \eqref{eq:estimate-sannino-shifman} denotes a real, positive constant at \(\mathcal{O}(1/N_{c})\).

Our lattice study published in \citep{DellaMorte:2023ylq} finds

\begin{equation}    
\dfrac{m_{\mathrm{P}}}{m_{\mathrm{S}}} = 0.356(54)\,.   
\end{equation}

The predictions from \citep{Sannino:2003xe} and \citep{Armoni:2005qr} are trivially compatible in the large-\(N_c\) limit but differ in the way they approach this limit. We now aim to test this result using a larger number of colours. This approach
reduces the \(\mathcal{O}(1/N_c^2)\) effects in the estimate \eqref{eq:estimate-sannino-shifman} and can therefore give an indication which of these two predictions, which are based on different sets of assumptions, describe the \(\mathcal{O}(1/N_c)\) effects more accurately.

In addition, we are tuning ensembles without clover-improvement in \(N_c=3\) to gauge how much the estimate at \(N_c=3\) from \citep{DellaMorte:2023ylq} is affected by cut-off effects.

\hypertarget{larger-n_c}{%
\section{\texorpdfstring{Larger \(N_c\)}{Larger N\_c}}\label{larger-n_c}}

For testing this prediction for \(SU(N_c)\)-theories with \(N_c>3\), we use \(\texttt{HiRep}\) \citep{DelDebbio:2008zf, hirep_repo} using again Symanzik improvement in the gauge action and one flavour of clover-improved Wilson fermions at \(c_{\mathrm{sw}} = 1\) in the two-index antisymmetric representation for \(N_c=4,5,6\). We have explored the parameter space for these three \(N_c\) to find approximate fake pion masses in the range of \(am_{\pi}^{\mathrm{fake}} \approx 0.2\) - \(1.0\) and lattice spacings around \(0.1~\mathrm{fm}\).

\hypertarget{matching-physical-parameters}{%
\subsection{Matching physical parameters}\label{matching-physical-parameters}}

For a reasonable estimate for the coupling \(\beta\) to yield lattice spacings of approximately \(0.1~\mathrm{fm}\), we are following the suggestion from t'Hooft in \citep{HOOFT1974461}, to keep the t'Hooft coupling \(\lambda = g^2 N_c\) constant. Since

\begin{equation}
\beta = \dfrac{N_c}{g^2} \Rightarrow \lambda = \dfrac{N_c^2}{\beta}\,,
\end{equation}

we can naively rescale

\begin{equation}
\beta \sim N_c^2\,,
\end{equation}

yielding the parameters displayed in table \ref{tab:betarescaling}.

In practice, the rescaling described in \citep{HOOFT1974461} features the renormalised coupling, and we initially expected the need to adjust \(\beta\) due to significant deviations from the target physical parameters. Surprisingly, we found good agreement in the resulting physical parameter space, yielding the same lattice spacings as for \(N_c=3\) by just using the naive scaling. Subsequently, it seems that the naive scaling suggested by t'Hooft works well in this setting, even if it is only a tree-level estimate.

Unfortunately, when naively scaling the \(\beta(N_c=3)=4.5\) to \(N_c=4,5\) and \(6\), we observed strong topological freezing in all simulations within the interesting physical parameter space. In the following, we detail ensembles to explore the parameter space to stay within target physical parameters but avoid topological freezing.

\begin{table}[ht]
\centering
\begin{tabular}{rrl}
  \hline
$N_c$ & $\beta_{\mathrm{naive}}$ & $\beta\approx$ \\ 
  \hline
3 & 4.50 & 4.5 \\ 
  4 & 8.00 & 7.6 \\ 
  5 & 12.50 & 12.1 \\ 
  6 & 18.00 & -- \\ 
   \hline
\end{tabular}
\caption{Naive estimates to rescale \(\beta\) in order to simulate at the right t'Hooft coupling, following \citep{HOOFT1974461} and an approximate value for \(\beta\) at which the freezing is sufficiently reduced.} 
\label{tab:betarescaling}
\end{table}

\hypertarget{parameter-dependence-of-topological-freezing}{%
\subsection{Parameter dependence of topological freezing}\label{parameter-dependence-of-topological-freezing}}

\begin{figure}

{\centering \includegraphics{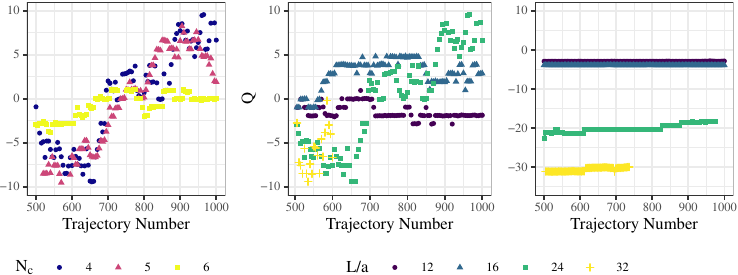} 

}

\caption{Severity of topological freezing depending on the parameter space. The left and middle image show pure gauge simulations over 500 trajectories depending on the spatial lattice extent at fixed \(N_c=4\) (middle) and depending on the number of colours at fixed \(L/a=24\) (left). The figure to the right shows the development of the topological charge for dynamic ensembles after thermalization for a fixed \(\kappa=0.1450\) depending on the spatial lattice extent \(L/a\). Note that the data for pure gauge was already published in \citep{Jaeger:2022ypq}.}\label{fig:top-freezing}
\end{figure}

For periodic boundary conditions, the parameter dependence of the number of topological sectors explored depending on volume and quark mass is derived in \citep{Leutwyler:1992yt}. For one-flavour theories, they derive that the probability of finding a gauge configuration in a given topological sector is normally distributed around \(Q=0\) as

\begin{equation}
\sigma = \mathrm{Var}(Q) = \langle Q^2\rangle = V\Sigma m
\end{equation}

where \(V\) is the volume \(T\times L^3\), \(m\) the quark mass and \(\Sigma\) describes a low energy constant related to the quark condensate in the chiral limit. Correspondingly, the probability of movement between different topological sectors is increasing with \(Vm\), so we can compensate a decrease in \(m\) with a corresponding increase in \(V\). The topological susceptibility will, in any case, go to zero in the chiral limit. The strength of the freezing may also depend on \(N_c\) and become worse when increasing \(N_c\). For this case we need to consider additional options to avoid autocorrelations of our ensembles.

Another stronger predictor for topological freezing is the lattice spacing, which we control over \(\beta\). Since \(\beta\) is inversely proportional to the lattice spacing, a slight change in \(\beta\) can have a significant effect on the lattice spacing; the autocorrelation time of the topological charge behaves as \(\tau_{Q}\sim a^{-5}\) for lattice spacings of \(\lesssim 0.1~\mathrm{fm}\) \citep{Luscher:2010we}. Relaxing \(\beta\) is a way to reduce topological freezing if this is sufficient for the precision one wants to achieve in the continuum limit.

\hypertarget{overview-of-simulation-parameters}{%
\subsection{Overview of simulation parameters}\label{overview-of-simulation-parameters}}

\begin{table}[ht]
\centering
\begin{tabular}{rrrrrlll}
  \hline
$N_c$ & $\beta$ & $\kappa$ & $L/a$ & $T/a$ & BCs & start & $Q$ \\ 
  \hline
3 & 4.50 & 0.1410 & 24 & 64 & periodic & hot & suff. \\ 
  3 & 4.50 & 0.1400 & 32 & 64 & periodic & hot & suff. \\ 
  4 & 8.00 & 0.1470 & 12 & 48 & periodic & hot & (-1) \\ 
  4 & 8.00 & 0.1450 & 24 & 48 & periodic & hot & (-20,-19,-18,-17) \\ 
  5 & 12.50 & 0.1520 & 16 & 48 & periodic & cold & (0) \\ 
  4 & 7.60 & 0.1520 & 16 & 48 & periodic & cold & suff. \\ 
  5 & 12.00 & 0.1530 & 16 & 48 & periodic & cold & (0,1,2) \\ 
  4 & 7.80 & 0.1440 & 16 & 96 & open & cold & suff. \\ 
  4 & 7.80 & 0.1440 & 16 & 192 & open & cold & suff. \\ 
   \hline
\end{tabular}
\caption{Selected ensembles for \(N_c>3\) to explore parameter space and technical study setup and two representative ensembles from the \(N_c=3\) study \citep{DellaMorte:2023ylq}. The column \emph{BCs} denotes the choice of boundary conditions, and the column \emph{start} the chosen start configuration using either a cold or a hot gauge configuration. All simulations were performed at \(c_{\mathrm{sw}}=1.0\) with trajectory length \(\tau=2.0\). The column \(Q\) indicates the topological sectors sampled after thermalisation, with \emph{suff.} indicating that there is no problem with topological freezing and that sufficiently many sectors were sampled.} 
\label{tab:higherncruns}
\end{table}

We see the effects described above in our ensembles. Table \ref{tab:higherncruns} shows a small selection of ensembles \(N_c > 3\) and two representative ensembles at \(N_c=3\) from \citep{DellaMorte:2023ylq} to illustrate the steps necessary to generate ergodic ensembles.

For \(N_c>3\), all ensembles are at approximately \(0.1~\mathrm{fm}\), and all ensembles are frozen in a single topological sector if the lattice is small. For these simulations, \(\beta\) is scaled naively. Increasing the lattice size can reduce the freezing, but using a hot start ensemble we may still only explore topological sectors far from zero, while a cold start allows thermalization near \(Q=0\).

\begin{figure}

{\centering \includegraphics{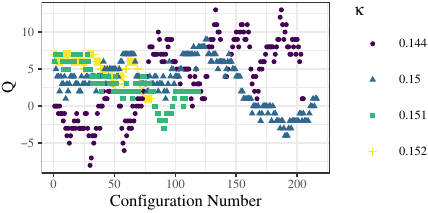} 

}

\caption{The topological charge moves after adjusting $\beta$ to $7.6$. We do see some signs of freezing again for light fake pion masses. Promisingly, $\kappa=0.1520$ corresponds to $am_{\pi}\approx 0.3$ and is therefore light enough for production. }\label{fig:tc-moving}
\end{figure}

We also generated ensembles with a slightly adjusted \(\beta\). As described in \citep{Luscher:2010we}, one can see that adjusting the spacing has the most significant effect on increasing the number of topological sectors explored. For the precision requirements of this study, it is sufficient to avoid lattices that are too fine for this technical reason and perform the continuum extrapolation from spacings that are not too strongly subjected to freezing. We found in our simulations that decreasing \(\beta\) from \(8.0\) to \(7.6\) completely solves the problem of topological freezing at sufficiently small fake pion masses and lattice spacings of around \(0.1~\mathrm{fm}\) at \(N_c=4\); see figure \ref{fig:tc-moving}.

\begin{figure}

{\centering \includegraphics{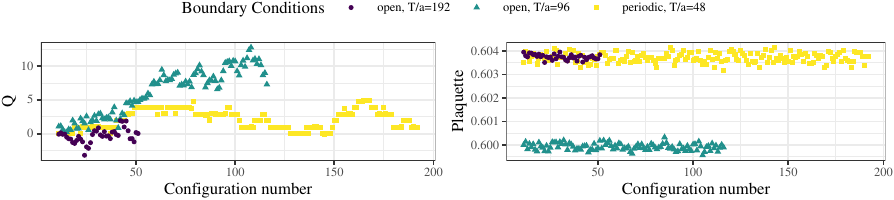} 

}

\caption{Finite-size effects for open boundary conditions. We see good agreement between periodic boundary conditions and open boundary conditions with T/a=192. The agreement is less good for T/a=96, however the effects are at subpercent level.}\label{fig:plaqopenbc}
\end{figure}

In case of topological freezing at light quark masses, for example, at \(N_c=5\) or \(6\) with fine lattice spacing, we have explored the option to use open boundary conditions in addition to large lattices. While open boundary conditions explore more topological sectors, we have to increase the temporal extent by a factor of two to measure sufficiently far away from the boundary. Our preliminary evaluations of the plaquette and the fake-pion mass show finite-size effects contributing at the per cent level. See figure \ref{fig:plaqopenbc} for comparing the plaquette between simulations with open and periodic boundary conditions.

\hypertarget{cut-off-effects}{%
\section{Cut-off effects}\label{cut-off-effects}}

\begin{table}[ht]
\centering
\begin{tabular}{rrrrrll}
  \hline
$N_c$ & $\beta$ & $\kappa$ & $L/a$ & $T/a$ & $am_{\pi}^{\mathrm{fake}}$ & $t^{\ast}$ \\ 
  \hline
3 & 4.555 & 0.1420 & 24 & 48 & 0.676(7) & 5.85(5) \\ 
  3 & 4.555 & 0.1450 & 24 & 48 & 0.508(7) & 6.37(5) \\ 
  3 & 4.555 & 0.1478 & 24 & 48 & 0.34(1) & 6.90(5) \\ 
   \hline
\end{tabular}
\caption{Overview of parameter choice for understanding cut-off effects. All ensembles are with periodic boundary conditions, hot start, $\tau=2.0$ and $c_{\mathrm{sw}}=0.0$. Results for $am_{\pi}^{\mathrm{fake}}$ and $t^{\ast}$ are preliminary.} 
\label{tab:runscutoff}
\end{table}

\begin{figure}

{\centering \includegraphics{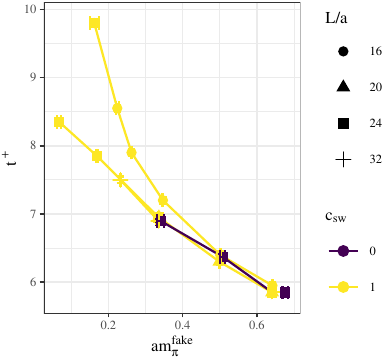} 

}

\caption{Finite-size effects in scale setting for \(N_c=3\), clover-improved data from \citep{DellaMorte:2023ylq}. \(t^{\ast}\) is defined as the reference scale for scale setting using the Wilson flow \citep{Luscher:2010iy}, using the symmetric definition of \(E\). We show preliminary results from ensembles at \(c_{\mathrm{sw}}=0\) with rescaled baseline \(\beta=4.555\).}\label{fig:fse-scalesetting}
\end{figure}

The deviations from the prediction found in the case of \(N_c=3\) might not only be due to effects of order \(1/N_c^2\) but can also arise due to cut-off effects. To understand the order of magnitude of cut-off effects, we are generating ensembles for \(N_c=3\) without clover-improvement that are tuned to achieve the same physical parameters as their clover-improved counterparts. We list planned ensembles with preliminary fake pion masses and \(t^{\ast}\) in table \ref{tab:runscutoff}.

\hypertarget{computational-cost}{%
\section{Computational cost}\label{computational-cost}}

For \(N_c=3\), the two-index antisymmetric representation coincides with the fundamental representation. This one-flavour QCD theory is supported well by different lattice community software libraries. Larger-\(N_c\) with Wilson fermions are supported by \texttt{HiRep} and \texttt{Grid} \citep{Boyle:2015tjk}, with some GPU-support in \texttt{Grid} and upcoming GPU-support for \texttt{HiRep}.

We compared benchmarks between \texttt{HiRep} and \texttt{Grid} for CPUs on LUMI-C and found that \texttt{HiRep} significantly outperformed \texttt{Grid} for Wilson fermions. However, the growing computational expense is unavoidable due to larger requirements for memory and compute. One can estimate that the necessary computational time scales at least with \(N_c^2\) due to the matrix multiplication in the Dirac operator. However, larger memory requirements degrade this scaling.

\hypertarget{conclusion}{%
\section{Conclusion}\label{conclusion}}

This is a progress report on the project. Moving from \(N_c=3\) to higher \(N_c\) presents several challenges. These challenges are a substantial growth of computational expense of the dynamical lattice calculations, severe topological freezing at sufficiently low masses and high coupling, and substantial finite-volume effects. We found that the right recipe for conducting a study of this kind at a relatively fine lattice spacing is to offset a decrease in quark mass with a proportional increase of the lattice volume, consider a possible increase in lattice spacing for ensembles that suffer from topological freezing, use GPUs to offset the significant computational and memory overhead, simulate at large lattices of, at the very least, \(L/a=20\) to control finite-size effects in scale setting, start from a unit configuration to thermalize around \(Q=0\), and, if needed, use open boundary conditions to avoid freezing additionally.

\hypertarget{acknowledgements}{%
\section{Acknowledgements}\label{acknowledgements}}

This project has received funding from the European Union's Horizon 2020 research and innovation program under the Marie Sk\l odowska-Curie grant agreements \textnumero 813942 and 894103. F.P.G.Z. acknowledges support from UKRI Future Leader Fellowship MR/T019956/1. We thank the EuroHPC Joint Undertaking for providing computational resources on the Discoverer supercomputer under proposal ID EHPC-REG-2023R01-102. Early simulations with \texttt{Grid} were performed on LUMI-C and LUMI-G hosted by CSC (Finland) and the LUMI consortium using resources provided by the University of Southern Denmark and the LUMI-G pilot program. Dynamic simulations with \texttt{HiRep} on GPUs were possible using GPU nodes provided by the UCloud interactive HPC system managed by the eScience Center at the University of Southern Denmark. We would like to thank Steffen Ulrik Jensen for his contributions to the exploration of the parameter space during his master thesis work.

\renewcommand\refname{References}
  \bibliography{literature.bib}

\begin{thebibliography}{15}%
\makeatletter
\providecommand \@ifxundefined [1]{%
 \@ifx{#1\undefined}
}%
\providecommand \@ifnum [1]{%
 \ifnum #1\expandafter \@firstoftwo
 \else \expandafter \@secondoftwo
 \fi
}%
\providecommand \@ifx [1]{%
 \ifx #1\expandafter \@firstoftwo
 \else \expandafter \@secondoftwo
 \fi
}%
\providecommand \natexlab [1]{#1}%
\providecommand \enquote  [1]{``#1''}%
\providecommand \bibnamefont  [1]{#1}%
\providecommand \bibfnamefont [1]{#1}%
\providecommand \citenamefont [1]{#1}%
\providecommand \href@noop [0]{\@secondoftwo}%
\providecommand \href [0]{\begingroup \@sanitize@url \@href}%
\providecommand \@href[1]{\@@startlink{#1}\@@href}%
\providecommand \@@href[1]{\endgroup#1\@@endlink}%
\providecommand \@sanitize@url [0]{\catcode `\\12\catcode `\$12\catcode
  `\&12\catcode `\#12\catcode `\^12\catcode `\_12\catcode `\%12\relax}%
\providecommand \@@startlink[1]{}%
\providecommand \@@endlink[0]{}%
\providecommand \url  [0]{\begingroup\@sanitize@url \@url }%
\providecommand \@url [1]{\endgroup\@href {#1}{\urlprefix }}%
\providecommand \urlprefix  [0]{URL }%
\providecommand \Eprint [0]{\href }%
\providecommand \doibase [0]{http://dx.doi.org/}%
\providecommand \selectlanguage [0]{\@gobble}%
\providecommand \bibinfo  [0]{\@secondoftwo}%
\providecommand \bibfield  [0]{\@secondoftwo}%
\providecommand \translation [1]{[#1]}%
\providecommand \BibitemOpen [0]{}%
\providecommand \bibitemStop [0]{}%
\providecommand \bibitemNoStop [0]{.\EOS\space}%
\providecommand \EOS [0]{\spacefactor3000\relax}%
\providecommand \BibitemShut  [1]{\csname bibitem#1\endcsname}%
\let\auto@bib@innerbib\@empty
\bibitem [{\citenamefont {Corrigan}\ and\ \citenamefont
  {Ramond}(1979)}]{CORRIGAN197973}%
  \BibitemOpen
  \bibfield  {author} {\bibinfo {author} {\bibfnamefont {E.}~\bibnamefont
  {Corrigan}}\ and\ \bibinfo {author} {\bibfnamefont {P.}~\bibnamefont
  {Ramond}},\ }\href {\doibase https://doi.org/10.1016/0370-2693(79)90022-4}
  {\bibfield  {journal} {\bibinfo  {journal} {Physics Letters B}\ }\textbf
  {\bibinfo {volume} {87}},\ \bibinfo {pages} {73} (\bibinfo {year}
  {1979})}\BibitemShut {NoStop}%
\bibitem [{\citenamefont {Sannino}\ and\ \citenamefont
  {Shifman}(2004)}]{Sannino:2003xe}%
  \BibitemOpen
  \bibfield  {author} {\bibinfo {author} {\bibfnamefont {F.}~\bibnamefont
  {Sannino}}\ and\ \bibinfo {author} {\bibfnamefont {M.}~\bibnamefont
  {Shifman}},\ }\href {\doibase 10.1103/PhysRevD.69.125004} {\bibfield
  {journal} {\bibinfo  {journal} {Phys. Rev. D}\ }\textbf {\bibinfo {volume}
  {69}},\ \bibinfo {pages} {125004} (\bibinfo {year} {2004})},\ \Eprint
  {http://arxiv.org/abs/hep-th/0309252} {arXiv:hep-th/0309252} \BibitemShut
  {NoStop}%
\bibitem [{\citenamefont {Armoni}\ and\ \citenamefont
  {Imeroni}(2005)}]{Armoni:2005qr}%
  \BibitemOpen
  \bibfield  {author} {\bibinfo {author} {\bibfnamefont {A.}~\bibnamefont
  {Armoni}}\ and\ \bibinfo {author} {\bibfnamefont {E.}~\bibnamefont
  {Imeroni}},\ }\href {\doibase 10.1016/j.physletb.2005.10.004} {\bibfield
  {journal} {\bibinfo  {journal} {Phys. Lett. B}\ }\textbf {\bibinfo {volume}
  {631}},\ \bibinfo {pages} {192} (\bibinfo {year} {2005})},\ \Eprint
  {http://arxiv.org/abs/hep-th/0508107} {arXiv:hep-th/0508107} \BibitemShut
  {NoStop}%
\bibitem [{\citenamefont {Della~Morte}\ \emph {et~al.}(2023)\citenamefont
  {Della~Morte}, \citenamefont {J\"ager}, \citenamefont {Sannino},
  \citenamefont {Tsang},\ and\ \citenamefont {Ziegler}}]{DellaMorte:2023ylq}%
  \BibitemOpen
  \bibfield  {author} {\bibinfo {author} {\bibfnamefont {M.}~\bibnamefont
  {Della~Morte}}, \bibinfo {author} {\bibfnamefont {B.}~\bibnamefont
  {J\"ager}}, \bibinfo {author} {\bibfnamefont {F.}~\bibnamefont {Sannino}},
  \bibinfo {author} {\bibfnamefont {J.~T.}\ \bibnamefont {Tsang}}, \ and\
  \bibinfo {author} {\bibfnamefont {F.~P.~G.}\ \bibnamefont {Ziegler}},\ }\href
  {\doibase 10.1103/PhysRevD.107.114506} {\bibfield  {journal} {\bibinfo
  {journal} {Phys. Rev. D}\ }\textbf {\bibinfo {volume} {107}},\ \bibinfo
  {pages} {114506} (\bibinfo {year} {2023})},\ \Eprint
  {http://arxiv.org/abs/2302.10514} {arXiv:2302.10514 [hep-lat]} \BibitemShut
  {NoStop}%
\bibitem [{\citenamefont {Lüscher}\ \emph {et~al.}()\citenamefont {Lüscher}
  \emph {et~al.}}]{openqcd}%
  \BibitemOpen
  \bibfield  {author} {\bibinfo {author} {\bibfnamefont {M.}~\bibnamefont
  {Lüscher}} \emph {et~al.},\ }\href@noop {} {\enquote {\bibinfo {title}
  {Openqcd},}\ }\bibinfo {howpublished}
  {\url{https://luscher.web.cern.ch/luscher/openQCD/}}\BibitemShut {NoStop}%
\bibitem [{\citenamefont {Clark}\ and\ \citenamefont
  {Kennedy}(2007)}]{Clark:2006fx}%
  \BibitemOpen
  \bibfield  {author} {\bibinfo {author} {\bibfnamefont {M.~A.}\ \bibnamefont
  {Clark}}\ and\ \bibinfo {author} {\bibfnamefont {A.~D.}\ \bibnamefont
  {Kennedy}},\ }\href {\doibase 10.1103/PhysRevLett.98.051601} {\bibfield
  {journal} {\bibinfo  {journal} {Phys. Rev. Lett.}\ }\textbf {\bibinfo
  {volume} {98}},\ \bibinfo {pages} {051601} (\bibinfo {year} {2007})},\
  \Eprint {http://arxiv.org/abs/hep-lat/0608015} {arXiv:hep-lat/0608015}
  \BibitemShut {NoStop}%
\bibitem [{\citenamefont {Francis}\ \emph {et~al.}(2018)\citenamefont
  {Francis}, \citenamefont {Hudspith}, \citenamefont {Lewis},\ and\
  \citenamefont {Tulin}}]{Francis:2018xjd}%
  \BibitemOpen
  \bibfield  {author} {\bibinfo {author} {\bibfnamefont {A.}~\bibnamefont
  {Francis}}, \bibinfo {author} {\bibfnamefont {R.~J.}\ \bibnamefont
  {Hudspith}}, \bibinfo {author} {\bibfnamefont {R.}~\bibnamefont {Lewis}}, \
  and\ \bibinfo {author} {\bibfnamefont {S.}~\bibnamefont {Tulin}},\ }\href
  {\doibase 10.1007/JHEP12(2018)118} {\bibfield  {journal} {\bibinfo  {journal}
  {JHEP}\ }\textbf {\bibinfo {volume} {12}},\ \bibinfo {pages} {118} (\bibinfo
  {year} {2018})},\ \Eprint {http://arxiv.org/abs/1809.09117} {arXiv:1809.09117
  [hep-ph]} \BibitemShut {NoStop}%
\bibitem [{\citenamefont {Del~Debbio}\ \emph {et~al.}(2010)\citenamefont
  {Del~Debbio}, \citenamefont {Patella},\ and\ \citenamefont
  {Pica}}]{DelDebbio:2008zf}%
  \BibitemOpen
  \bibfield  {author} {\bibinfo {author} {\bibfnamefont {L.}~\bibnamefont
  {Del~Debbio}}, \bibinfo {author} {\bibfnamefont {A.}~\bibnamefont {Patella}},
  \ and\ \bibinfo {author} {\bibfnamefont {C.}~\bibnamefont {Pica}},\ }\href
  {\doibase 10.1103/PhysRevD.81.094503} {\bibfield  {journal} {\bibinfo
  {journal} {Phys. Rev. D}\ }\textbf {\bibinfo {volume} {81}},\ \bibinfo
  {pages} {094503} (\bibinfo {year} {2010})},\ \Eprint
  {http://arxiv.org/abs/0805.2058} {arXiv:0805.2058 [hep-lat]} \BibitemShut
  {NoStop}%
\bibitem [{\citenamefont {Pica}\ \emph {et~al.}()\citenamefont {Pica} \emph
  {et~al.}}]{hirep_repo}%
  \BibitemOpen
  \bibfield  {author} {\bibinfo {author} {\bibfnamefont {C.}~\bibnamefont
  {Pica}} \emph {et~al.},\ }\href@noop {} {\enquote {\bibinfo {title}
  {Hirep},}\ }\bibinfo {howpublished}
  {\url{https://github.com/claudiopica/HiRep}}\BibitemShut {NoStop}%
\bibitem [{\citenamefont {{'t Hooft}}(1974)}]{HOOFT1974461}%
  \BibitemOpen
  \bibfield  {author} {\bibinfo {author} {\bibfnamefont {G.}~\bibnamefont {{'t
  Hooft}}},\ }\href {\doibase https://doi.org/10.1016/0550-3213(74)90154-0}
  {\bibfield  {journal} {\bibinfo  {journal} {Nuclear Physics B}\ }\textbf
  {\bibinfo {volume} {72}},\ \bibinfo {pages} {461} (\bibinfo {year}
  {1974})}\BibitemShut {NoStop}%
\bibitem [{\citenamefont {J{\"a}ger}\ \emph {et~al.}(2023)\citenamefont
  {J{\"a}ger}, \citenamefont {Della~Morte}, \citenamefont {Martins},
  \citenamefont {Sannino}, \citenamefont {Tsang},\ and\ \citenamefont
  {Ziegler}}]{Jaeger:2022ypq}%
  \BibitemOpen
  \bibfield  {author} {\bibinfo {author} {\bibfnamefont {B.}~\bibnamefont
  {J{\"a}ger}}, \bibinfo {author} {\bibfnamefont {M.}~\bibnamefont
  {Della~Morte}}, \bibinfo {author} {\bibfnamefont {S.}~\bibnamefont
  {Martins}}, \bibinfo {author} {\bibfnamefont {F.}~\bibnamefont {Sannino}},
  \bibinfo {author} {\bibfnamefont {J.~T.}\ \bibnamefont {Tsang}}, \ and\
  \bibinfo {author} {\bibfnamefont {F.~P.~G.}\ \bibnamefont {Ziegler}},\ }\href
  {\doibase 10.22323/1.430.0212} {\bibfield  {journal} {\bibinfo  {journal}
  {PoS}\ }\textbf {\bibinfo {volume} {LATTICE2022}},\ \bibinfo {pages} {212}
  (\bibinfo {year} {2023})},\ \Eprint {http://arxiv.org/abs/2212.06709}
  {arXiv:2212.06709 [hep-lat]} \BibitemShut {NoStop}%
\bibitem [{\citenamefont {Leutwyler}\ and\ \citenamefont
  {Smilga}(1992)}]{Leutwyler:1992yt}%
  \BibitemOpen
  \bibfield  {author} {\bibinfo {author} {\bibfnamefont {H.}~\bibnamefont
  {Leutwyler}}\ and\ \bibinfo {author} {\bibfnamefont {A.~V.}\ \bibnamefont
  {Smilga}},\ }\href {\doibase 10.1103/PhysRevD.46.5607} {\bibfield  {journal}
  {\bibinfo  {journal} {Phys. Rev. D}\ }\textbf {\bibinfo {volume} {46}},\
  \bibinfo {pages} {5607} (\bibinfo {year} {1992})}\BibitemShut {NoStop}%
\bibitem [{\citenamefont {L{\"u}scher}(2010)}]{Luscher:2010we}%
  \BibitemOpen
  \bibfield  {author} {\bibinfo {author} {\bibfnamefont {M.}~\bibnamefont
  {L{\"u}scher}},\ }\href {\doibase 10.22323/1.105.0015} {\bibfield  {journal}
  {\bibinfo  {journal} {PoS}\ }\textbf {\bibinfo {volume} {LATTICE2010}},\
  \bibinfo {pages} {015} (\bibinfo {year} {2010})},\ \Eprint
  {http://arxiv.org/abs/1009.5877} {arXiv:1009.5877 [hep-lat]} \BibitemShut
  {NoStop}%
\bibitem [{\citenamefont {L\"uscher}(2010)}]{Luscher:2010iy}%
  \BibitemOpen
  \bibfield  {author} {\bibinfo {author} {\bibfnamefont {M.}~\bibnamefont
  {L\"uscher}},\ }\href {\doibase 10.1007/JHEP08(2010)071} {\bibfield
  {journal} {\bibinfo  {journal} {JHEP}\ }\textbf {\bibinfo {volume} {08}},\
  \bibinfo {pages} {071} (\bibinfo {year} {2010})},\ \bibinfo {note} {[Erratum:
  JHEP 03, 092 (2014)]},\ \Eprint {http://arxiv.org/abs/1006.4518}
  {arXiv:1006.4518 [hep-lat]} \BibitemShut {NoStop}%
\bibitem [{\citenamefont {Boyle}\ \emph {et~al.}(2015)\citenamefont {Boyle},
  \citenamefont {Yamaguchi}, \citenamefont {Cossu},\ and\ \citenamefont
  {Portelli}}]{Boyle:2015tjk}%
  \BibitemOpen
  \bibfield  {author} {\bibinfo {author} {\bibfnamefont {P.}~\bibnamefont
  {Boyle}}, \bibinfo {author} {\bibfnamefont {A.}~\bibnamefont {Yamaguchi}},
  \bibinfo {author} {\bibfnamefont {G.}~\bibnamefont {Cossu}}, \ and\ \bibinfo
  {author} {\bibfnamefont {A.}~\bibnamefont {Portelli}},\ }\href@noop {} {\
  (\bibinfo {year} {2015})},\ \Eprint {http://arxiv.org/abs/1512.03487}
  {arXiv:1512.03487 [hep-lat]} \BibitemShut {NoStop}%
\end{thebibliography}%

\end{document}